\documentclass[12pt]{article}
\usepackage{amsmath}
\usepackage{graphicx}
\usepackage{float}
\begin{document}

\title{{\bf Are Most Particles Gravitons?}
\thanks{Alberta-Thy-24-16, arXiv:YYMM.NNNNN [hep-th], honorable mention in the 2016 Gravity Research Foundation Awards for Essays on Gravitation}}

\author{
Don N. Page
\thanks{Internet address:
profdonpage@gmail.com}
\\
Theoretical Physics Institute\\
Department of Physics\\
4-183 CCIS\\
University of Alberta\\
Edmonton, Alberta T6G 2E1\\
Canada
}

\date{2016 March 31}

\maketitle
\large
\begin{abstract}
\baselineskip 22 pt

The number of baryons in the observable universe is of the order of $10^{80}$, as is the number of electrons.  The number of photons is about nine orders of magnitude greater, $10^{89}$, as is the estimated number of neutrinos.  However, the number of gravitons could be more than twenty orders of magnitude larger yet, of the order of $10^{113}\, r$, where $r$ is the tensor-to-scalar ratio for quantum fluctuations produced by inflation, which could be as high as $0.1$. 

\end{abstract}

\baselineskip 22 pt

\newpage

It is not known whether our universe has a finite or infinite number of particles.  However, the observable universe, the part that any one observer can see, has a finite number.  This number is much larger than the number of particles in a human or snowflake, though much smaller than the exponentials of the latter numbers.  (This is why no two humans or snowflakes are observed to be exactly alike.)

A mnemonic spatially flat $\Lambda$CDM Friedmann-Lema{\^\i}tre-Robertson-Walker metric \cite{Scott:2013oib} for the large-scale universe after roughly the time that it became transparent is
\begin{equation}
ds^2 = - dt^2 + a^2(t)(dx^2 + dy^2 + dz^2),
\label{FLRW}
\end{equation}
where the cosmic scale factor is
\begin{equation}
a(t) = \{(2/3)\sinh[(3/2)H_\infty t]\}^{2/3},
\label{scale}
\end{equation}
where $\Lambda = 3 H_\infty^2$ is within $1\,\sigma$ of ten square attohertz ($10^{-35}$ s$^{-2}$), of $(10\,{\rm Gyr})^{-2}$, and of $3\pi/(5^3 2^{400})$ in Planck units ($\hbar = c = G = 4\pi\epsilon_0 = k_{\rm Boltzmann} = 1$), this last coincidence being equivalent to the fact that the Gibbons-Hawking entropy of the de Sitter spacetime that our universe seems to be headed for is $S_{\rm dS} = 3\pi/\Lambda \approx 5^3 2^{400}$.  Note that the coefficient of the scale factor has been set so that $a \sim (H_\infty t)^{2/3}$ during nonrelativistic matter domination when $H_\infty t \ll 1$.  The mnemonic coincidence \cite{Scott:2013oib} for our temporal location in this spacetime is that at the present time, $t = t_0$, one has $a_0 \equiv a(t_0) \approx 1$, within $1\,\sigma$.  We can then derive the age $t_0 \approx (2/\sqrt{3\Lambda})\ln{(1.5+\sqrt{3.25})} \approx 4.36\times 10^{17}\, {\rm s} \approx 13.80\, {\rm Gyr} \approx 8.07\times 10^{60}$ Planck times and the present Hubble constant $H_0 \approx (\sqrt{13}/3) H_\infty = \sqrt{13\Lambda/27} \approx 1.19\times 10^{-61} \approx 2.19\ {\rm aHz\  (attohertz},\ 10^{-18}\ {\rm hertz}) \approx 67.85\ $km/s/Mpc.

We can also deduce from this mnemonic that the ratio of the dark energy density to the critical density today ($\rho_{\rm crit} = 3H_0^2/(8\pi G) \approx 13\Lambda/(72\pi) = 13/(24\,S_{\rm dS}) \approx (13/3000)\, 2^{-400}$ in Planck units) is $\Omega_\Lambda \approx 9/13$, and that the ratio of the nonrelativistic matter density to the critical density is $\Omega_{\rm m} \approx 4/13$.  The additional mnemonic coincidence \cite{Scott:2013oib} for the corresponding ratios $\Omega_{\rm b}$ for the baryon density and $\Omega_{\rm c}$ for the cold dark matter density (with $\Omega_{\rm b} + \Omega_{\rm c} = \Omega_{\rm m}$), namely
$\Omega_{\rm c}/\Omega_{\rm b} \approx 2\Omega_\Lambda/\Omega_{\rm c}$, allows us further to deduce $\Omega_{\rm b} \approx (13-3\sqrt{17})/13$ and $\Omega_{\rm c} \approx (3\sqrt{17}-9)/13$.

Yet another mnemonic coincidence is that the square of the ratio of the Planck mass, $M_{\rm Pl} = \sqrt{\hbar c/G}$, to the proton mass, $m_{\rm p}$, is within half of one per cent of the largest prime ever found by humans without the use of computers, $(M_{\rm Pl}/m_{\rm p})^2 \approx 2^{127}-1$.  From this and from $\rho_{\rm crit}$ and $\Omega_{\rm b}$, one can deduce the approximation that the current average cosmological number density of baryons in Planck units is $n_{\rm b} \approx 1.06\times 10^{-105}$.

Similarly, the mnemonic coincidence \cite{Scott:2013oib} that the present cosmic microwave background (CMB) radiation temperature is approximately one hundred octaves, eight perfect fifths, and one justly tuned minor fifth below the Planck temperature, or $T_\gamma \approx (160/3^8)2^{-100}$, allows one to deduce that the current cosmological number density of photons is $n_\gamma \approx 2^{-184}\, 3^{-24}\, 5^3\, \pi^{-2}\, \zeta(3) \approx 1.734\times 10^{-96}$ in Planck units.  The number density of neutrinos is expected to be $9/11$ as large, or $n_\nu \approx 1.419\times 10^{-96}$, and the total number density of photons and neutrinos is expected to be about $3.153\times 10^{-96}$ in Planck units, or about 747 particles per cubic centimeter jetting across the universe.

To get the total numbers of these particles within the observable universe, we need to calculate how large it is.  The present distance to the furthest electromagnetic structures that one can see (the CMB) works out to be very nearly
\begin{align}
R &\approx 3\cdot 12^{1/6}\, \Lambda^{-1/2}
  \approx 18^{1/3}\, 5^{3/2}\, \pi^{-1/2}\, 2^{200}
  \nonumber \\
  &\approx 2.65\times 10^{61}
  \approx 30\cdot12^{1/6}\ {\rm Gyr}
  \approx 45.4\ {\rm Gyr}.
\label{R}
\end{align}
Then the volume of the present (spatially flat) universe out to this radius (the ``observable'' universe, though of course this is really what has evolved from what we can only see as it existed in the past) is, in Planck units,
\begin{equation}
V_0 = (4\pi/3) R^3
    \approx 24\cdot 5^{9/2}\, \pi^{-1/2}\, 2^{600}
    \approx 7.85\times 10^{184}.
\label{V_0}
\end{equation}

Multiplying this volume by the number densities of baryons, photons, and neutrinos gives their total numbers within the observable universe under these mnemonic approximations as
\begin{align}
&N_{\rm b} \approx 2^{527/2}\, 5^{3/2}\, (13-3\sqrt{17})\, \pi^{-1/2}
          \approx 8.34\times 10^{79},  
\label{N_b}\\
&N_\gamma \approx 2^{319}\, 3^{-23}\, 5^{15/2}\, \pi^{-5/2}\,\zeta(3)
         \approx 1.36\times 10^{89},
\label{N_gamma}\\
&N_\nu \approx 2^{319}\, 3^{-21}\, 5^{15/2}\, 11^{-1}\, \pi^{-5/2}\,\zeta(3)
      \approx 1.11\times 10^{89},
\label{N_nu}\\
&N_{\rm part} \approx 2^{321}\, 3^{-23}\, 5^{17/2}\, 11^{-1}\, \pi^{-5/2}\,\zeta(3)
      \approx 2.48\times 10^{89},
\label{N_part}
\end{align}
where the last quantity is the estimated total number in the observable universe of all known particles other than gravitons.

Now let us try to estimate the number of gravitons in the observable universe.  The analogue of the cosmic microwave background for gravitons would have a present temperature several times cooler than the photons (because the photons decoupled much later and hence had many other sources of energy dumped into it) and hence a number density more than an order of magnitude lower.  However, inflation is predicted to have produced a large number of gravitons by quantum fluctuations \cite{Starobinsky:1979ty, Rubakov:1982df, Fabbri:1983us, Abbott:1984fp, Starobinsky:1985ww}.  The number density is higher at lower frequencies, so let us look at the lowest possible frequency for a gravitational wave to have one wavelength totally within the observable universe of diameter $2R$, which gives a minimum angular frequency
\begin{equation}
\omega_{\rm m} = \pi/R \approx 1.18\times 10^{-61}
\label{omega_m}
\end{equation}
in Planck units, or a frequency in cycles per second of
\begin{equation}
f_{\rm m} = \omega_{\rm m}/(2\pi)
 = 1/(2R) \approx 0.349\ {\rm aHz}. 
\label{f_m}
\end{equation}

The total number density at the present time of gravitons from primordial slow-roll inflation, with this minimum frequency, works out \cite{Maggiore:1999vm, Boyle:2005se} to be roughly (ignoring transfer functions that should be fairly close to unity for the low frequencies that dominate)
\begin{equation}
n_{\rm g} \sim \frac{H_*^2\, H_0^2}{2\pi^2\, \omega_{\rm m}} \sim \frac{H_*^2\, H_0^2\, R}{2\pi^3}.
\label{n_g}
\end{equation}
Here $H_*$ is the Hubble expansion rate when the mode with present frequency $\omega_{\rm m}$ grew larger than the Hubble size $H_*^{-1}$ at that time.  Then the total number of gravitons within what the observable universe evolves to on a hypersurface of today's constant cosmic time is
\begin{equation}
N_{\rm g} = n_{\rm g} V_0 \sim \frac{2}{3\pi^2}\, H_*^2\, H_0^2\, R^4.
\label{N_g1}
\end{equation}

Using the formulas above for the Hubble constant $H_0$ and the present radius $R$ of the observable universe, the formula $S_{\rm dS} = 3\pi/\Lambda \approx 5^3 2^{400}$ for the late-time Gibbons-Hawking entropy, and the inflaton potential energy density $V_* = 3H_*^2/(8\pi)$, one can also write the total number of gravitons as
\begin{equation}
N_{\rm g} \sim \frac{208\cdot 12^{2/3}}{9\pi^2}\, S_{\rm dS}\, V_*
    \sim 12.3\, S_{\rm dS}\, V_* \sim 3.96\times 10^{123}\, V_*.
\label{N_g2}
\end{equation}
That is, the total number of gravitons produced by in inflation within the observable universe is of the order of the Gibbons-Hawking entropy multiplied by the inflaton potential energy density in Planck units.

Alternatively, if one uses the asymptotic value of the energy density of the dark matter represented by the cosmological constant, $\rho_\infty = \Lambda/(8\pi)$, then one can express the total number of gravitons as
\begin{equation}
N_{\rm g} \sim \frac{26\cdot 12^{2/3}}{3\pi^2} \frac{V_*}{\rho_\infty}
    \sim 4.60\, \frac{V_*}{\rho_\infty} \sim 6.65\, \frac{V_*}{\rho_{\rm crit}},
\label{N_g3}
\end{equation}
which is of the order of the ratio of the energy density during inflation to the asymptotic energy density or to the critical energy density today.  The ratio $V_*/\rho_\infty$ is also the ratio of the entropies of de Sitter spaces with the late-time value of the cosmological constant and with the inflationary value of the cosmological `constant' respectively.

The value of the inflationary potential energy density $V_*$ can be written in terms of the amplitude $A_{\rm s}$ of scalar perturbations from slow-roll inflation and of the tensor-to-scalar ratio $r$ and is given by the {\it Planck} collaboration \cite {Ade:2015lrj} as
\begin{equation}
V_* = \frac{3 A_{\rm s} r}{128},
\label{V}
\end{equation}
where in my Planck units ($\hbar = c = G = 1$) the reduced Planck mass given in that paper is $M_{pl} = (8\pi G/\hbar c)^{-1/2} = (8\pi)^{-1/2}$.

Using the value in the last column of Table 4 on page 31 of that same paper \cite{Ade:2015lrj}, $A_{\rm s} \approx 10^{-10} \exp{(3.064)} \approx 2.14\times 10^{-9} \approx 2^5\,3^{-14}\,5^{-5}$, one then gets
\begin{equation}
N_{\rm g} \sim 2.0\times 10^{113}\, r \sim 8.0\times 10^{23}\, r\, N_{\rm part}.
\label{N_g4}
\end{equation}

The value of the tensor-to-scalar ratio $r$ is not yet known, as there is as of yet no clear evidence for tensor perturbations from inflation and hence for a nonzero $r$, and an upper limit \cite{Ade:2015tva, Ade:2015lrj} is $r < 0.08$ at the 95\% confidence level.  However, since slow-roll inflation with a value of $V_*$ not too many orders of magnitude below the Planck density seems highly plausible, it does appear highly likely that there are far more gravitons in the observable universe than the total for all other particles.  Even a value of $r$ as low as $0.000125 = 1/8000$, which would not be measurable in the foreseeable future, would give about twenty orders of magnitude more gravitons in the observable universe than the total for all other known particles.

One should note that although the total number of gravitons in the observable universe may be only several orders of magnitude less than the asymptotic de Sitter entropy, nearly all of these gravitons are in coherent states, so the entropy they contribute is far less than their numbers.  Most of the entropy of the observable universe could be in the cosmic microwave background radiation of photons and neutrinos (or in the Bekenstein-Hawking coarse-grained entropies of supermassive black holes, though their von Neumann entropies at this very early stage of their Hawking evaporation would be expected to be far less than the von Neumann entropy in the CMB).

In summary, the total number of gravitons in the observable universe at the present time, produced by inflation, seems likely to be far more than the total for all other particles combined.  The ratio is predicted to be comparable to the ratio of the energy density during inflation to the energy density today or in the asymptotic future.  Truly ours appears to be a gravitational universe.

This work was supported in part by the Natural Sciences and Engineering Research Council of Canada.

\end{document}